\documentclass[twocolumn,pre,superscriptaddress]{revtex4}

\usepackage{dcolumn}
\usepackage{amsmath}

\usepackage{graphicx}
\usepackage{rotating}
\usepackage{multirow}

\begin{document}

% Macros
\renewcommand{\d}{\mathrm{d}}
\renewcommand{\O}{\mathrm{O}}
\newcommand{\e}{\mathrm{e}}
\newcommand{\half}{\mbox{$\frac12$}}
\newcommand{\set}[1]{\lbrace#1\rbrace}
\newcommand{\av}[1]{\langle#1\rangle}
\newcommand{\eref}[1]{(\ref{#1})}
\newcommand{\etal}{{\it{}et~al.}}
\newcommand{\defn}{\textit}
\newcommand{\from}{\leftarrow}
\newcommand{\ve}{\mathbf{e}}
\newcommand{\vI}{\mathbf{I}}
\newcommand{\vM}{\mathbf{M}}
\newcommand{\vV}{\mathbf{V}}
\newcommand{\vD}{\mathbf{D}}
\newcommand{\vA}{\mathbf{A}}
\newcommand{\vT}{\mathbf{T}}
\newcommand{\vS}{\mathbf{S}}
\newcommand{\vs}{\mathbf{s}}
\newcommand{\vX}{\mathbf{X}}
\newcommand{\vx}{\mathbf{x}}
\newcommand{\vone}{\mathbf{1}}
\newcommand{\Tr}{\mathop{\rm Tr}}
\newcommand{\norm}[1]{\left\|\,#1\,\right\|}

% Style parameters
\newlength{\figurewidth}
\ifdim\columnwidth<10.5cm
  \setlength{\figurewidth}{0.95\columnwidth}
\else
  \setlength{\figurewidth}{10cm}
\fi
\setlength{\parskip}{0pt}
\setlength{\tabcolsep}{6pt}
\setlength{\arraycolsep}{2pt}

\title{Finding and evaluating community structure in networks}
\author{M. E. J. Newman}
\affiliation{Department of Physics and Center for the Study of Complex
Systems,\\
University of Michigan, Ann Arbor, MI 48109--1120}
\affiliation{Santa Fe Institute, 1399 Hyde Park Road, Santa Fe, NM 87501}
\author{M. Girvan}
\affiliation{Santa Fe Institute, 1399 Hyde Park Road, Santa Fe, NM 87501}
\affiliation{Department of Physics, Cornell University, Ithaca,
NY 14853--2501}

\begin{abstract}
We propose and study a set of algorithms for discovering community
structure in networks---natural divisions of network nodes into densely
connected subgroups.  Our algorithms all share two definitive features:
first, they involve iterative removal of edges from the network to split it
into communities, the edges removed being identified using one of a number
of possible ``betweenness'' measures, and second, these measures are,
crucially, recalculated after each removal.  We also propose a measure for
the strength of the community structure found by our algorithms, which
gives us an objective metric for choosing the number of communities into
which a network should be divided.  We demonstrate that our algorithms are
highly effective at discovering community structure in both
computer-generated and real-world network data, and show how they can be
used to shed light on the sometimes dauntingly complex structure of
networked systems.
\end{abstract}
\maketitle

\section{Introduction}
\begin{figure}[b]
\begin{center}
\resizebox{7.5cm}{!}{\includegraphics{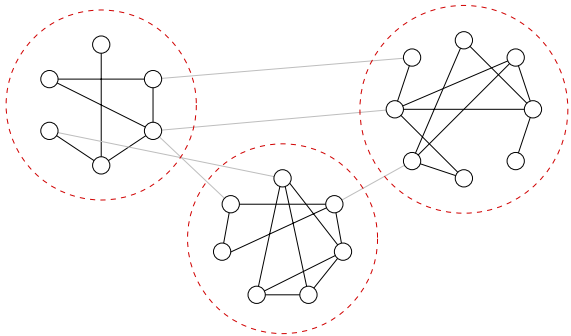}}
\end{center}
\caption{A small network with community structure of the type considered in
this paper.  In this case there are three communities, denoted by the
dashed circles, which have dense internal links but between which there are
only a lower density of external links.}
\label{example}
\end{figure}

Empirical studies and theoretical modeling of networks have been the
subject of a large body of recent research in statistical physics and
applied mathematics~\cite{Strogatz01,AB02,DM03b,Newman03d}.  Network ideas
have been applied with great success to topics as diverse as the Internet
and the world wide web~\cite{FFF99,AJB99,Broder00},
epidemiology~\cite{KG99,MN00a,PV01a,ML01}, scientific citation and
collaboration~\cite{Redner98,Newman01a}, metabolism~\cite{Jeong00,WF01},
and ecosystems~\cite{DWM02a,CGA02}, to name but a few.  A property that
seems to be common to many networks is \defn{community structure}, the
division of network nodes into groups within which the network connections
are dense, but between which they are sparser---see Fig.~\ref{example}.
The ability to find and analyze such groups can provide invaluable help in
understanding and visualizing the structure of networks.  In this paper we
show how this can be achieved.

The study of community structure in networks has a long history.  It is
closely related to the ideas of graph partitioning in graph theory and
computer science, and hierarchical clustering in
sociology~\cite{GJ79,Scott00}.  Before presenting our own findings, it is
worth reviewing some of this preceding work, to understand its
achievements and where it falls short.

Graph partitioning is a problem that arises in, for example, parallel
computing.  Suppose we have a number~$n$ of intercommunicating computer
processes, which we wish to distribute over a number~$g$ of computer
processors.  Processes do not necessarily need to communicate with all
others, and the pattern of required communications can be represented by a
graph or network in which the vertices represent processes and edges join
process pairs that need to communicate.  The problem is to allocate the
processes to processors in such a way as roughly to balance the load on
each processor, while at the same time minimizing the number of edges that
run between processors, so that the amount of interprocessor communication
(which is normally slow) is minimized.  In general, finding an exact
solution to a partitioning task of this kind is believed to be an
NP-complete problem, making it prohibitively difficult to solve for large
graphs, but a wide variety of heuristic algorithms have been developed that
give acceptably good solutions in many cases, the best known being perhaps
the Kernighan--Lin algorithm~\cite{KL70}, which runs in time $\O(n^3)$ on
sparse graphs.

A solution to the graph partitioning problem is however not particularly
helpful for analyzing and understanding networks in general.  If we merely
want to find if and how a given network breaks down into communities, we
probably don't know how many such communities there are going to be, and
there is no reason why they should be roughly the same size.  Furthermore,
the number of inter-community edges needn't be strictly minimized either,
since more such edges are admissible between large communities than between
small ones.

\begin{figure}
\begin{center}
\resizebox{6cm}{!}{\includegraphics{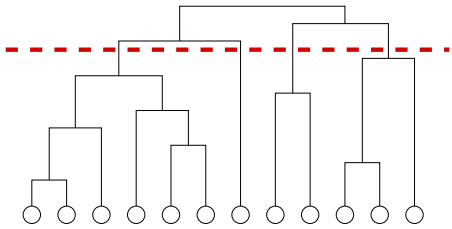}}
\end{center}
\caption{A hierarchical tree or dendrogram illustrating the type of output
generated by the algorithms described here.  The circles at the bottom of
the figure represent the individual vertices of the network.  As we move up
the tree the vertices join together to form larger and larger communities,
as indicated by the lines, until we reach the top, where all are joined
together in a single community.  Alternatively, we the dendrogram depicts
an initially connected network splitting into smaller and smaller
communities as we go from top to bottom.  A cross-section of the tree at
any level, as indicated by the dotted line, will give the communities at
that level.  The vertical height of the split-points in the tree are
indicative only of the order in which the splits (or joins) took place,
although it is possible to construct more elaborate dendrograms in which
these heights contain other information.}
\label{dendrogram}
\end{figure}

As far as our goals in this paper are concerned, a more useful approach is
that taken by social network analysis with the set of techniques known as
hierarchical clustering.  These techniques are aimed at discovering natural
divisions of (social) networks into groups, based on various metrics of
similarity or strength of connection between vertices.  They fall into two
broad classes, agglomerative and divisive~\cite{Scott00}, depending on
whether they focus on the addition or removal of edges to or from the
network.  In an agglomerative method, similarities are calculated by one
method or another between vertex pairs, and edges are then added to an
initially empty network ($n$~vertices with no edges) starting with the
vertex pairs with highest similarity.  The procedure can be halted at any
point, and the resulting components in the network are taken to be the
communities.  Alternatively, the entire progression of the algorithm from
empty graph to complete graph can be represented in the form of a tree or
\defn{dendrogram} such as that shown in Fig.~\ref{dendrogram}.  Horizontal
cuts through the tree represent the communities appropriate to different
halting points.

Agglomerative methods based on a wide variety of similarity measures have
been applied to different networks.  Some networks have natural similarity
metrics built in.  For example, in the widely studied network of
collaborations between film actors~\cite{WS98,ASBS00}, in which two actors
are connected if they have appeared in the same film, one could quantify
similarity by how many films actors have appeared in together~\cite{ML00}.
Other networks have no natural metric, but suitable ones can be devised
using correlation coefficients, path lengths, or matrix methods.  A well
known example of an agglomerative clustering method is the Concor algorithm
of Breiger~\etal~\cite{BBA75}.

\begin{figure}
\begin{center}
\resizebox{\figurewidth}{!}{\includegraphics{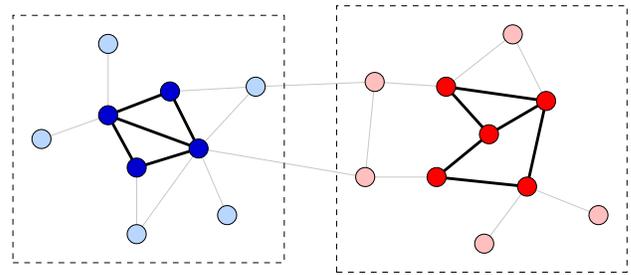}}
\end{center}
\caption{Agglomerative clustering methods are typically good at discovering
the strongly linked cores of communities (bold vertices and edges) but tend
to leave out peripheral vertices, even when, as here, most of them clearly
belong to one community or another.}
\label{problem}
\end{figure}

Agglomerative methods have their problems however.  One concern is that
they fail with some frequency to find the correct communities in networks
were the community structure is known, which makes it difficult to place
much trust in them in other cases.  Another is their tendency to find only
the cores of communities and leave out the periphery.  The core nodes in a
community often have strong similarity, and hence are connected early in
the agglomerative process, but peripheral nodes that have no strong
similarity to others tend to get neglected, leading to structures like that
shown in Fig.~\ref{problem}.  In this figure, there are a number of
peripheral nodes whose community membership is obvious to the eye---in most
cases they have only a single link to a specific community---but
agglomerative methods often fail to place such nodes correctly.

In this paper, therefore, we focus on divisive methods.  These methods have
been relatively little studied in the previous literature, either in social
network theory or elsewhere, but, as we will see, seem to offer a lot of
promise.  In a divisive method, we start with the network of interest and
attempt to find the \emph{least} similar connected pairs of vertices and
then remove the edges between them.  By doing this repeatedly, we divide
the network into smaller and smaller components, and again we can stop the
process at any stage and take the components at that stage to be the
network communities.  Again, the process can be represented as a dendrogram
depicting the successive splits of the network into smaller and smaller
groups.

The approach we take follows roughly these lines, but adopts a somewhat
different philosophical viewpoint.  Rather than looking for the most weakly
connected vertex pairs, our approach will be to look for the edges in the
network that are most ``between'' other vertices, meaning that the edge is,
in some sense, responsible for connecting many pairs of others.  Such edges
need not be weak at all in the similarity sense.  How this idea works out
in practice will become clear in the course of the presentation.

Briefly then, the outline of this paper is as follows.  In
Sec.~\ref{algorithms} we describe the crucial concepts behind our methods
for finding community structure in networks and show how these concepts can
be turned into a concrete prescription for performing calculations.  In
Sec.~\ref{implementation} we describe in detail the implementation of our
methods.  In Sec.~\ref{gos} we consider ways of determining when a
particular division of a network into communities is a good one, allowing
us to quantify the success of our community-finding algorithms.  And in
Sec.~\ref{apps} we give a number of applications of our algorithms to
particular networks, both real and artificial.  In Sec.~\ref{concs} we give
our conclusions.  A brief report of some of the work contained in this
paper has appeared previously as Ref.~\onlinecite{GN02}.

\section{Finding communities in a network}
\label{algorithms}
In this paper we present a class of new algorithms for network clustering,
i.e.,~the discovery of community structure in networks.  Our discussion
focuses primarily on networks with only a single type of vertex and a
single type of undirected, unweighted edge, although generalizations to
more complicated network types are certainly possible.

There are two central features that distinguish our algorithms from those
that have preceded them.  First, our algorithms are divisive rather than
agglomerative.  Divisive algorithms have occasionally been studied in the
past, but, as discussed in the introduction, ours differ in focusing not on
removing the edges between vertex pairs with lowest similarity, but on
finding edges with the highest ``betweenness,'' where betweenness is some
measure that favors edges that lie between communities and disfavors those
that lie inside communities.

To make things more concrete, we give some examples of the types of
betweenness measures we will be looking at.  All of them are based on the
same idea.  If two communities are joined by only a few inter-community
edges, then all paths through the network from vertices in one community to
vertices in the other must pass along one of those few edges.  Given a
suitable set of paths, one can count how many go along each edge in the
graph, and this number we then expect to be largest for the inter-community
edges, thus providing a method for identifying them.  Our different
measures correspond to various implementations of this idea.
\begin{enumerate}
\item The simplest example of such a betweenness measure is that based on
shortest (geodesic) paths: we find the shortest paths between all pairs of
vertices and count how many run along each edge.  To the best of our
knowledge this measure was first introduced by Anthonisse in a
never-published technical report in 1971~\cite{Anthonisse71}.  Anthonisse
called it ``rush,'' but we prefer the term \defn{edge betweenness}, since
the quantity is a natural generalization to edges of the well-known
(vertex) betweenness measure of Freeman~\cite{Freeman77}, which was the
inspiration for our approach.  When we need to distinguish it from the
other betweenness measures considered in this paper, we will refer to it as
\defn{shortest-path betweenness}.  A fast algorithm for calculating the
shortest-path betweenness is given in Sec.~\ref{eb}.
\item The shortest-path betweenness can be thought of in terms of signals
traveling through a network.  If signals travel from source to destination
along geodesic network paths, and all vertices send signals at the same
constant rate to all others, then the betweenness is a measure of the rate
at which signals pass along each edge.  Suppose however that signals do not
travel along geodesic paths, but instead just perform a random walk about
the network until they reach their destination.  This gives us another
measure on edges, the \defn{random-walk betweenness}: we calculate the
expected net number of times that a random walk between a particular pair
of vertices will pass down a particular edge and sum over all vertex pairs.
The random-walk betweenness can be calculated using matrix methods, as
described in Sec.~\ref{rw}.
\item Another betweenness measure is motivated by ideas from elementary
circuit theory.  We consider the circuit created by placing a unit
resistance on each edge of the network and unit current source and sink at
a particular pair of vertices.  The resulting current flow in the network
will travel from source to sink along a multitude of paths, those with
least resistance carrying the greatest fraction of the current.  The
\defn{current-flow betweenness} for an edge we define to be the absolute
value of the current along the edge summed over all source/sink pairs.  It
can be calculated using Kirchhoff's laws, as described in
Sec.~\ref{resistors}.  In fact, as we will show, the current-flow
betweenness turns out to be exactly the same as the random walk betweenness
of the previous paragraph, but we nonetheless consider it separately since
it leads to a simpler derivation of the measure.
\end{enumerate}

These measures are only suggestions; many others are possible and may well
be appropriate for specific applications.  Measures (1) and~(2) are in some
sense extremes in the spectrum of possibilities, one corresponding to
signals that know exactly where they are going, and the other to signals
that have no idea where they are going.  As we will see, however, these two
measures actually give rather similar results, indicating that the precise
choice of betweenness measure may not, at least for the types of
applications considered here, be that important.

The second way in which our methods differ from previous ones is in the
inclusion of a ``recalculation step'' in the algorithm.  If we were to
perform a standard divisive clustering based on edge betweenness we would
calculate the edge betweenness for all edges in the network and then remove
edges in decreasing order of betweenness to produce a dendrogram like that
of Fig.~\ref{dendrogram}, showing the order in which the network split up.

However, once the first edge in the network is removed in such an
algorithm, the betweenness values for the remaining edges will no longer
reflect the network as it now is.  This can give rise to unwanted
behaviors.  For example, if two communities are joined by two edges, but,
for one reason or another, most paths between the two flow along just one
of those edges, then that edge will have a high betweenness score and the
other will not.  An algorithm that calculated betweennesses only once and
then removed edges in betweenness order would remove the first edge early
in the course of its operation, but the second might not get removed until
much later.  Thus the obvious division of the network into two parts might
not be discovered by the algorithm.  In the worst case the two parts
themselves might be individually broken up before the division between the
two is made.  In practice, problems like this crop up in real networks with
some regularity and render algorithms of this type ineffective for the
discovery of community structure.

The solution, luckily, is obvious.  We simply recalculate our betweenness
measure after the removal of each edge.  This certainly adds to the
computational effort of performing the calculation, but its effect on the
results is so desirable that we consider the price worth paying.

Thus the general form of our community structure finding algorithm is as
follows:
\begin{enumerate}
\item Calculate betweenness scores for all edges in the network.
\item Find the edge with the highest score and remove it from the network.
\item Recalculate betweenness for all remaining edges.
\item Repeat from step 2.
\end{enumerate}

In fact, it appears that the recalculation step is the most important
feature of the algorithm, as far as getting satisfactory results is
concerned.  As mentioned above, our studies indicate that, once one hits on
the idea of using betweenness measures to weight edges, the exact measure
one uses appears not to influence the results highly.  The recalculation
step, on the other hand, is absolutely crucial to the operation of our
methods.  This step was missing from previous attempts at solving the
clustering problem using divisive algorithms, and yet without it the
results are very poor indeed, failing to find known community structure
even in the simplest of cases.  In Sec.~\ref{karate} we give an example
comparing the performance of the algorithm on a particular network with and
without the recalculation step.

In the following sections we discuss implementation and give examples of
our algorithms for finding community structure.  For the reader who merely
wants to know what algorithm they should use for their own problem, let us
give an immediate answer: for most problems, we recommend the algorithm
with betweenness scores calculated using the shortest-path betweenness
measure (1) above.  This measure appears to work well and is the quickest
to calculate---as described in Sec.~\ref{eb}, it can be calculated for all
edges in time $\O(mn)$, where $m$ is the number of edges in the graph and
$n$ is the number of vertices.  This is the only version of the algorithm
that we discussed in Ref.~\onlinecite{GN02}~\footnote{Following the
publication of that paper, the algorithm has been implemented in the
commercial network analysis software package UCInet and in the open-source
network library JUNG.  (See \texttt{http://www.analytictech.com/} and
\texttt{http://jung.sourceforge.net/} respectively.)}.  The other versions
we discuss, while being of some pedagogical interest, make greater
computational demands, and in practice seem to give results no better than
the shortest-path method.

\section{Implementation}
\label{implementation}
In theory, the descriptions of the last section completely define the
methods we consider in this paper, but in practice there are a number of
tricks to their implementation that are important for turning the
description into a workable computer algorithm.

Essentially all of the work in the algorithm is in the calculation of the
betweenness scores for the edges; the job of finding and removing the
highest-scoring edge is trivial and not computationally demanding.  Let us
tackle our three suggested betweenness measures in turn.

\subsection{Shortest-path betweenness}
\label{eb}
At first sight, it appears that calculating the edge betweenness measure
based on geodesic paths for all edges will take $\O(mn^2)$ operations on a
graph with $m$ edges and $n$ vertices: calculating the shortest path
between a particular pair of vertices can be done using breadth-first
search in time $\O(m)$~\cite{AMO93,CLRS01}, and there are $\O(n^2)$ vertex
pairs.  Recently however new algorithms have been proposed by
Newman~\cite{Newman01c} and independently by Brandes~\cite{Brandes01} that
can perform the calculation faster than this, finding all betweennesses in
$\O(mn)$ time.  Both Newman and Brandes gave algorithms for the standard
Freeman vertex betweenness, but it is trivial to adapt their algorithms for
edge betweenness.  We describe the resulting method here for the algorithm
of Newman.

\begin{figure}
\begin{center}
\resizebox{\columnwidth}{!}{\includegraphics{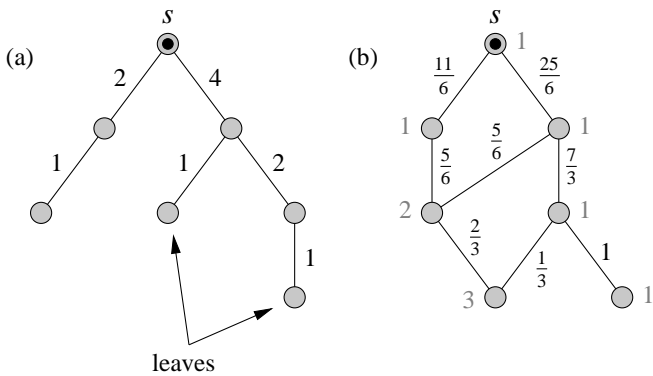}}
\end{center}
\caption{Calculation of shortest-path betweenness: (a)~When there is only
a single shortest path from a source vertex~$s$ (top) to all other
reachable vertices, those paths necessarily form a tree, which makes the
calculation of the contribution to betweenness from this set of paths
particularly simple, as describe in the text.  (b)~For cases in which there
is more than one shortest path to some vertices, the calculation is more
complex.  First we must calculate the number of paths from the source to
each other vertex (numbers on vertices), and then these are used to weight
the path counts appropriately.  In either case, we can check the results by
confirming that the sum of the betweennesses of the edges connected to the
source vertex is equal to the total number of reachable vertices---six in
each of the cases illustrated here.}
\label{betweenness}
\end{figure}

Breadth-first search can find shortest paths from a single vertex~$s$ to
all others in time~$\O(m)$.  In the simplest case, when there is only a
single shortest path from the source vertex to any other (we will consider
other cases in a moment) the resulting set of paths forms a shortest-path
tree---see Fig.~\ref{betweenness}a.  We can now use this tree to calculate
the contribution to betweenness for each edge from this set of paths as
follows.  We find first the ``leaves'' of the tree, i.e.,~those nodes such
that no shortest paths to other nodes pass through them, and we assign a
score of~1 to the single edge that connects each to the rest of the tree,
as shown in the figure.  Then, starting with those edges that are farthest
from the source vertex on the tree, i.e.,~lowest in
Fig.~\ref{betweenness}a, we work upwards, assigning a score to each edge
that is 1~plus the sum of the scores on the neighboring edges immediately
below it.  When we have gone though all edges in the tree, the resulting
scores are the betweenness counts for the paths from vertex~$s$.  Repeating
the process for all possible vertices~$s$ and summing the scores, we arrive
at the full betweenness scores for shortest paths between all pairs.  The
breadth-first search and the process of working up through the tree both
take worst-case time $\O(m)$ and there are $n$ vertices total, so the
entire calculation takes time $\O(mn)$ as claimed.

This simple case serves to illustrate the basic principle behind the
algorithm.  In general, however, it is not the case that there is only a
single shortest path between any pair of vertices.  Most networks have at
least some vertex pairs between which there are several geodesic paths of
equal length.  Figure~\ref{betweenness}b shows a simple example of a
shortest path ``tree'' for a network with this property.  The resulting
structure is in fact no longer a tree, and in such cases an extra step is
required in the algorithm to calculate the betweenness correctly.

In the traditional definition of vertex betweenness~\cite{Freeman77}
multiple shortest paths between a pair of vertices are given equal weights
summing to~1.  For example, if there are three shortest paths, each will be
given weight~$\frac13$.  We adopt the same definition for our edge
betweenness (as did Anthonisse in his original work~\cite{Anthonisse71},
although other definitions are possible~\cite{GKK01b}).  Note that the
paths may run along the same edge or edges for some part of their length,
resulting in edges with greater weight.  To calculate correctly what
fraction of the paths flow along each edge in the network, we generalize
the breadth-first search part of the calculation, as follows.

Consider Fig.~\ref{betweenness}b and suppose we are performing a
breadth-first search starting at vertex~$s$.  We carry out the following
steps:
\begin{enumerate}
\item The initial vertex~$s$ is given distance $d_s=0$ and a
weight~$w_s=1$.
\item Every vertex~$i$ adjacent to $s$ is given distance $d_i=d_s+1=1$, and
weight $w_i=w_s=1$.
\item For each vertex~$j$ adjacent to one of \emph{those} vertices~$i$ we do
one of three things:
  \begin{enumerate}
  \item If $j$ has not yet been assigned a distance, it is assigned
    distance $d_j=d_i+1$ and weight $w_j=w_i$.
  \item If $j$ has already been assigned a distance and $d_j=d_i+1$, then the
    vertex's weight is increased by~$w_i$, that is $w_j\from w_j+w_i$.
  \item If $j$ has already been assigned a distance and $d_j<d_i+1$, we do
    nothing.
  \end{enumerate}
\item Repeat from step 3 until no vertices remain that have assigned
distances but whose neighbors do not have assigned distances.
\end{enumerate}
In practice, this algorithm can be implemented most efficiently using a
queue or first-in/first-out buffer to store the vertices that have been
assigned a distance, just as in the standard breadth-first search.

Physically, the weight on a vertex~$i$ represents the number of distinct
paths from the source vertex to~$i$.  These weights are precisely what we
need to calculate our edge betweennesses, because if two vertices $i$
and~$j$ are connected, with $j$ farther than~$i$ from the source~$s$, then
the fraction of a geodesic path from $j$ through $i$ to $s$ is given by
$w_i/w_j$.  Thus, to calculate the contribution to edge betweenness from
all shortest paths starting at~$s$, we need only carry out the following
steps:
\begin{enumerate}
\item Find every ``leaf'' vertex~$t$, i.e.,~a vertex such that no paths
from $s$ to other vertices go though~$t$.
\item For each vertex~$i$ neighboring~$t$ assign a score to the edge from
$t$ to $i$ of $w_i/w_t$.
\item Now, starting with the edges that are farthest from the source
vertex~$s$---lower down in a diagram such as Fig.~\ref{betweenness}b---work
up towards~$s$.  To the edge from vertex~$i$ to vertex~$j$, with $j$ being
farther from $s$ than~$i$, assign a score that is 1~plus the sum of the
scores on the neighboring edges immediately below it (i.e.,~those with
which it shares a common vertex), all multiplied by $w_i/w_j$.
\item Repeat from step 3 until vertex~$s$ is reached.
\end{enumerate}
Now repeating this process for all~$n$ source vertices~$s$ and summing the
resulting scores on the edges gives us the total betweenness for all edges
in time $\O(mn)$.

We now have to repeat this calculation for each edge removed from the
network, of which there are $m$, and hence the complete community structure
algorithm based on shortest-path betweenness operates in worst-case time
$\O(m^2n)$, or $\O(n^3)$ time on a sparse graph.  In our experience, this
typically makes it tractable for networks of up to about $n=10\,000$
vertices, with current (\textit{circa} 2003) desktop computers.  In some
special cases one can do better.  In particular, we note that the removal
of an edge only affects the betweenness of other edges that fall in the
same component, and hence that we need only recalculate betweennesses in
that component.  Networks with strong community structure often break apart
into separate components quite early in the progress of the algorithm,
substantially reducing the amount of work that needs to be done on
subsequent steps.  Whether this results in a change in the computational
complexity of the algorithm for any commonly occurring classes of graphs is
an open question, but it certainly gives a substantial speed boost to many
of the calculations described in this paper.

Some networks are directed, i.e.,~their edges run in one direction only.
The world wide web is an example; links in the web point in one direction
only from one web page to another.  One could imagine a generalization of
the shortest-path betweenness that allowed for directed edges by counting
only those paths that travel in the forward direction along edges.  Such a
calculation is a trivial variation on the one described above.  However, we
have found that in many cases it is better to ignore the directed nature of
a network in calculating community structure.  Often an edge acts simply as
an indication of a connection between two nodes, and its direction is
unimportant.  For example, in Ref.~\onlinecite{GN02} we applied our
algorithm to a food web of predator-prey interactions between marine
species.  Predator-prey interactions are clearly directed---one species may
eat another, but it is unlikely that the reverse is simultaneously true.
However, as far as community structure is concerned, we want to know only
which species have interactions with which others.  We find, therefore,
that our algorithm applied to the undirected version of the food web works
well at picking out the community structure, and no special algorithm is
needed for the directed case.  We give another example of our method
applied to a directed graph in Sec.~\ref{other}.

\subsection{Resistor networks}
\label{resistors}
As examples of betweenness measures that take more than just shortest paths
into account, we proposed in Sec.~\ref{algorithms} measures based on random
walks and resistor networks.  In fact, as we now show, when appropriately
defined these two measures are precisely the same.  Here we derive the
resistance measure first, since it turns out to be simpler; in the
following section we derive the random walk measure and show that the two
are equivalent.

\begin{figure}
\begin{center}
\resizebox{\figurewidth}{!}{\includegraphics{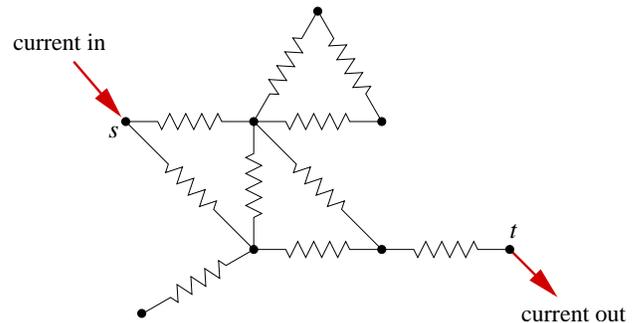}}
\end{center}
\caption{An example of the type of resistor network considered here, in
which a unit resistance is placed on each edge and unit current flows into
and out of the source and sink vertices.}
\label{rnet}
\end{figure}

Consider the network created by placing a unit resistance on every edge of
our network, a unit current source at vertex~$s$, and a unit current sink
at vertex~$t$ (see Fig.~\ref{rnet}).  Clearly the current between $s$ and
$t$ will flow primarily along short paths, but some will flow along longer
ones, roughly in inverse proportion to their length.  We will use the
absolute magnitude of the current flow as our betweenness score for each
source/sink pair.

The current flows in the network are governed by Kirchhoff's laws.  To
solve them we proceed as follows for each separate component of the graph.
Let $V_i$ be the voltage at vertex~$i$, measured relative to any convenient
point.  Then for all~$i$ we have
\begin{equation}
\sum_j A_{ij} (V_i-V_j) = \delta_{is} - \delta_{it},
\label{kirchhoff}
\end{equation}
where $A_{ij}$ is the $ij$ element of the adjacency matrix of the graph,
i.e.,~$A_{ij}=1$ if $i$ and $j$ are connected by an edge and $A_{ij}=0$
otherwise.  The left-hand side of Eq.~\eref{kirchhoff} represents the net
current flow out of vertex~$i$ along edges of the network, and the
right-hand side represents the source and sink.  Defining $k_i=\sum_j
A_{ij}$, which is the vertex degree, and creating a diagonal matrix~$\vD$
with these degrees on the diagonal $D_{ii}=k_i$, this equation can be
written in matrix form as $(\vD-\vA)\cdot\vV=\vs$, where the source vector
$\vs$ has components
\begin{equation}
s_i = \Biggl\lbrace\begin{array}{rl}
        +1 & \qquad\mbox{for $i=s$}\\
        -1 & \qquad\mbox{for $i=t$}\\
        0  & \qquad\mbox{otherwise.}
      \end{array}
\end{equation}

We cannot directly invert the matrix $\vD-\vA$ to get the voltage
vector~$\vV$, because the matrix (which is just the graph Laplacian) is
singular.  This is equivalent to saying that there is one undetermined
degree of freedom corresponding to the choice of reference potential for
measuring the voltages.  We can add any constant to a solution for the
vertex voltages and get another solution---only the voltage differences
matter.  In choosing the reference potential, we fix this degree of
freedom, leaving only $n-1$ more to be determined.  In mathematical terms,
once any $n-1$ of the equations in our matrix formulation are satisfied,
the remaining one is also automatically satisfied so long as current is
conserved in the network as a whole, i.e.,~so long as $\sum_i s_i=0$, which
is clearly true in this case.

Choosing any vertex~$v$ to be the reference point, therefore, we remove the
row and column corresponding to that vertex from $\vD$ and $\vA$ before
inverting.  Denoting the resulting $(n-1)\times(n-1)$ matrices $\vD_v$
and~$\vA_v$, we can then write
\begin{equation}
\vV = (\vD_v-\vA_v)^{-1}\cdot\vs.
\end{equation}

Calculation of the currents in the network thus involves inverting
$\vD_v-\vA_v$ once for any convenient choice of~$v$, and taking the
differences of pairs of columns to get the voltage vector $\vV$ for each
possible source/sink pair.  (The voltage for the one missing vertex~$v$ is
always zero, by hypothesis.)  The absolute magnitudes of the differences of
voltages along each edge give us betweenness scores for the given source
and sink.  Summing over all sources and sinks, we then get our complete
betweenness score.

The matrix inversion takes time $\O(n^3)$ in the worst case, while the
subsequent calculation of betweennesses takes time $\O(mn^2)$, where as
before $m$ is the number of edges and $n$ the number of vertices in the
graph.  Thus, the entire community structure algorithm, including the
recalculation step, will take $\O\bigl((n+m)mn^2\bigr)$ time to complete,
or $\O(n^4)$ on a sparse graph.  Although, as we will see, the algorithm is
good at finding community structure, this poor performance makes it
practical only for smaller graphs; a few hundreds of vertices is the most
that we have been able to do.  It is for this reason that we recommend
using the shortest-path betweenness algorithm in most cases, which gives
results about as good or better with considerably less effort.

\subsection{Random walks}
\label{rw}
The random-walk betweenness described in Sec.~\ref{algorithms} requires us
to calculate how often on average random walks starting at vertex~$s$ will
pass down a particular edge from vertex~$v$ to vertex~$w$ (or \textit{vice
versa}) before finding their way to a given target vertex~$t$.  To
calculate this quantity we proceed as follows for each separate component
of the graph.

As before, let $A_{ij}$ be an element of the adjacency matrix such that
$A_{ij}=1$ if vertices $i$ and $j$ are connected by an edge and $A_{ij}=0$
otherwise.  Consider a random walk that on each step decides uniformly
between the neighbors of the current vertex~$j$ and takes a step to one of
them.  The number of neighbors is just the degree of the vertex $k_j=\sum_i
A_{ij}$, and the probability for the transition from $j$ to $i$ is
$A_{ij}/k_j$, which we can regard as an element of the matrix
$\vM=\vA\cdot\vD^{-1}$, where $\vD$ is the diagonal matrix with
$D_{ii}=k_i$.

We are interested in walks that terminate when they reach the target~$t$,
so that $t$ is an absorbing state.  The most convenient way to represent
this is just to remove entirely the vertex~$t$ from the graph, so that no
walk ever reaches any other vertex from~$t$.  Thus let
$\vM_t=\vA_t\cdot\vD_t^{-1}$ be the matrix~$\vM$ with the $t$th row and
column removed (and similarly for $\vA_t$ and~$\vD_t$).

Now the probability that a walk starts at~$s$, takes $n$ steps, and ends up
at some other vertex (not~$t$), is given by the $is$ element of~$\vM_t^n$,
which we denote $[\vM_t^n]_{is}$.  In particular, walks end up at $v$ and
$w$ with probabilities $[\vM_t^n]_{vs}$ and $[\vM_t^n]_{ws}$, and of those
a fraction $1/k_v$ and $1/k_w$ respectively then pass along the edge
$(v,w)$ in one direction or the other.  (Note that they may also have
passed along this edge an arbitrary number of times before reaching this
point.)  Summing over all~$n$, the mean number of times that a walk of any
length traverses the edge from $v$ to $w$ is $k_v^{-1}
[(\vI-\vM_t)^{-1}]_{vs}$, and similarly for walks that go from $w$ to~$v$.

To highlight the similarity with the current-flow betweenness of
Sec.~\ref{resistors}, let us denote these two numbers $V_v$ and~$V_w$
respectively.  Then we can write
\begin{equation}
\vV = \vD^{-1}\cdot(\vI-\vM_t)^{-1}\cdot\vs
    = (\vD_t - \vA_t)^{-1}\cdot\vs,
\label{defsvv}
\end{equation}
where the source vector~$\vs$ is the vector whose components are all~0
except for a single~1 in the position corresponding to the source
vertex~$s$.

Now we define our random-walk betweenness for the edge $(v,w)$ to be the
absolute value of the \emph{difference} of the two probabilities $V_v$
and~$V_w$, i.e.,~the net number of times the walk passes along the edge in
one direction.  This seems a natural definition---it makes little sense to
accord an edge high betweenness simply because a walk went back and forth
along it many times.  It is the difference between the numbers of times the
edge is traversed in either direction that matters~\footnote{In fact we
have tried counting each traversal separately, but this method gives
extremely poor results, confirming our intuition that this would not be a
good betweenness measure.}.

But now we see that this method is very similar to the resistor network
calculation of Sec.~\ref{resistors}.  In that calculation we also evaluated
$(\vD_t-\vA_t)^{-1}\cdot\vs$ for a suitable source vector and then took
differences of the resulting numbers.  The only difference is that in the
current-flow calculation we had a sink term in~$\vs$ as well as a source.
Purely for the purposes of mathematical convenience, we can add such a sink
in the present case at the target vertex~$t$---this makes no difference to
the solution for $\vV$ since the $t$th row has been removed from the
equations anyway.  By doing this, however, we turn the equations into
precisely the form of the current-flow calculation, and hence it becomes
clear that the two measures are numerically identical, although their
derivation is quite different.  (It also immediately follows that we can
remove any row or column and still get the same answer---it doesn't have to
be row and column~$t$, although physically this choice makes the most
sense.)

\section{Quantifying the strength of community structure}
\label{gos}
As we show in Sec.~\ref{apps}, our community structure algorithms do an
excellent job of recovering known communities both in artificially
generated random networks and in real-world examples.  However, in
practical situations the algorithms will normally be used on networks for
which the communities are not known ahead of time.  This raises a new
problem: how do we know when the communities found by the algorithm are
good ones?  Our algorithms always produce \emph{some} division of the
network into communities, even in completely random networks that have no
meaningful community structure, so it would be useful to have some way of
saying how good the structure found is.  Furthermore, the algorithms'
output is in the form of a dendrogram which represents an entire nested
hierarchy of possible community divisions for the network.  We would like
to know which of these divisions are the best ones for a given
network---where we should cut the dendrogram to get a sensible division of
the network.

To answer these questions we now define a measure of the quality of a
particular division of a network, which we call the \defn{modularity}.
This measure is based on a previous measure of assortative mixing proposed
by Newman~\cite{Newman03c}.  Consider a particular division of a network
into $k$ communities.  Let us define a $k\times k$ symmetric matrix~$\ve$
whose element $e_{ij}$ is the fraction of all edges in the network that
link vertices in community~$i$ to vertices in community~$j$~\footnote{As
discussed in~\cite{Newman03c}, it is crucial to make sure each edge is
counted only once in the matrix $e_{ij}$---the same edge should not appear
both above and below the diagonal.  Alternatively, an edge linking
communities $i$ and~$j$ can be split, half-and-half, between the $ij$ and
$ji$ elements, which has the advantage of making the matrix symmetric.
Either way, there are a number of factors of~2 in the calculation that must
be watched carefully, lest they escape one's attention and make mischief.}.
(Here we consider all edges in the original network---even after edges have
been removed by the community structure algorithm our modularity measure is
calculated using the full network.)

The trace of this matrix $\Tr\ve=\sum_i e_{ii}$ gives the fraction of edges
in the network that connect vertices in the same community, and clearly a
good division into communities should have a high value of this trace.  The
trace on its own, however, is not a good indicator of the quality of the
division since, for example, placing all vertices in a single community
would give the maximal value of $\Tr\ve=1$ while giving no information
about community structure at all.

So we further define the row (or column) sums $a_i=\sum_j e_{ij}$, which
represent the fraction of edges that connect to vertices in community~$i$.
In a network in which edges fall between vertices without regard for the
communities they belong to, we would have $e_{ij}=a_i a_j$.  Thus we can
define a modularity measure by
\begin{equation}
Q = \sum_i \bigl( e_{ii} - a_i^2 \bigr) = \Tr\ve - \norm{\ve^2},
\label{defsq}
\end{equation}
where $\norm{\vx}$ indicates the sum of the elements of the matrix~$\vx$.
This quantity measures the fraction of the edges in the network that
connect vertices of the same type (i.e.,~within-community edges) minus the
expected value of the same quantity in a network with the same community
divisions but random connections between the vertices.  If the number of
within-community edges is no better than random, we will get $Q=0$.  Values
approaching $Q=1$, which is the maximum, indicate strong community
structure~\footnote{In Ref.~\onlinecite{Newman03c}, the measure was
normalized by dividing by its value on a network with perfect mixing, so
that we always get~1 for such a network.  We find however that doing this
in the present case masks some of the useful information to be gained from
the value of~$Q$, and hence that it is better to use the unnormalized
measure.  In general, this unnormalized measure will not reach a value
of~1, even on a perfectly mixed network.}.  In practice, values for such
networks typically fall in the range from about $0.3$ to~$0.7$.  Higher
values are rare.

The expected error on $Q$ can be calculated by treating each edge in the
network as an independent measurement of the contributions to the elements
of the matrix~$\ve$.  A simple jackknife procedure works
well~\cite{Efron79,Newman03c}.

Typically, we will calculate $Q$ for each split of a network into
communities as we move down the dendrogram, and look for local peaks in its
value, which indicate particularly satisfactory splits.  Usually we find
that there are only one or two such peaks and, as we will show in the next
section, in cases where the community structure is known beforehand by some
means we find that the positions of these peaks correspond closely to the
expected divisions.  The height of a peak is a measure of the strength of
the community division.

\section{Applications}
\label{apps}
In this section we give a number of applications of our algorithms to
particular problems, illustrating their operation, and their use in
understanding the structure of complex networks.

\subsection{Tests on computer-generated networks}
First, as a controlled test of how well our algorithms perform, we have
generated networks with known community structure, to see if the algorithms
can recognize and extract this structure.

\begin{figure}
\begin{center}
\resizebox{\figurewidth}{!}{\includegraphics[angle=90]{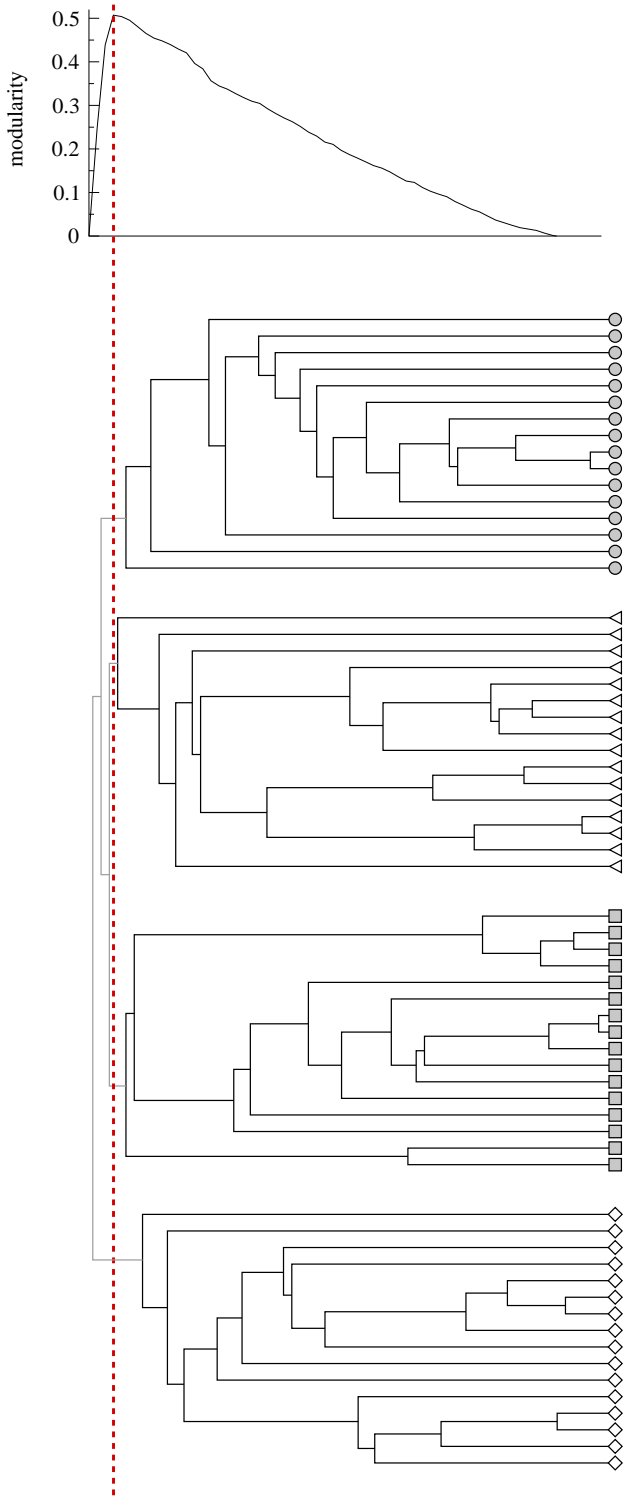}}
\end{center}
\caption{Plot of the modularity and dendrogram for a 64-vertex random
community-structured graph generated as described in the text with, in this
case, $z_\mathrm{in}=6$ and $z_\mathrm{out}=2$.  The shapes on the right
denote the four communities in the graph and as we can see, the peak in the
modularity (dotted line) corresponds to a perfect identification of the
communities.}
\label{rgtree}
\end{figure}

We have generated a large number of graphs with $n=128$ vertices, divided
into four communities of 32 vertices each.  Edges were placed independently
at random between vertex pairs with probability $p_\mathrm{in}$ for an edge
to fall between vertices in the same community and $p_\mathrm{out}$ to fall
between vertices in different communities.  The values of $p_\mathrm{in}$
and $p_\mathrm{out}$ were chosen to make the expected degree of each vertex
equal to~16.  In Fig.~\ref{rgtree} we show a typical dendrogram from the
analysis of such a graph using the shortest-path betweenness version of our
algorithm.  (In fact, for the sake of clarity, the figure is for a 64-node
version of the graph.)  Results for the random walk version are similar.
At the top of the figure we also show the modularity, Eq.~\eref{defsq}, for
the same calculation, plotted as a function of position in the dendrogram.
That is, the graph is aligned with the dendrogram so that one can read off
modularity values for different divisions of the network directly.  As we
can see, the modularity has a single clear peak at the point where the
network breaks into four communities, as we would expect.  The peak value
is around $0.5$, which is typical.

\begin{figure}
\begin{center}
\resizebox{\columnwidth}{!}{\includegraphics{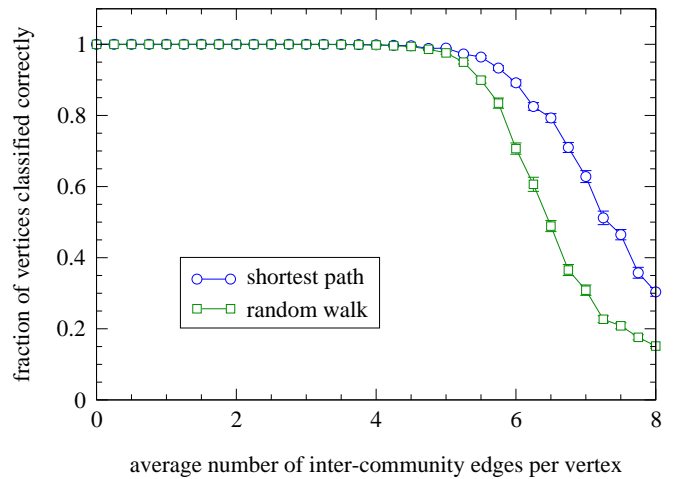}}
\end{center}
\caption{The fraction of vertices correctly identified by our algorithms in
the computer-generated graphs described in the text.  The two curves show
results for the edge betweenness (circles) and random walk (squares)
versions of the algorithm as a function of the number of edges vertices
have to others outside their own community.  The point $z_\mathrm{out}=8$
at the rightmost edge of the plot represents the point at which the
graphs---in this example---have as many connections outside their own
community as inside it.  Each point is an average over 100 graphs.}
\label{correct}
\end{figure}

In Fig.~\ref{correct} we show the fraction of vertices in our
computer-generated network sample classified correctly into the four
communities by our algorithms, as a function of the mean number
$z_\mathrm{out}$ of edges from each vertex to vertices in other
communities.  As the figure shows, both the shortest-path and random-walk
versions of the algorithm perform excellently, with more than 90\% of all
vertices classified correctly from $z_\mathrm{out}=0$ all the way to around
$z_\mathrm{out}=6$.  Only for $z_\mathrm{out}\gtrsim6$ does the
classification begin to deteriorate markedly.  In other words, our
algorithm correctly identifies the community structure in the network
almost all the way to the point $z_\mathrm{out}=8$ at which each vertex has
on average the same number of connections to vertices outside its community
as it does to those inside.

The shortest-path version of the algorithm does however perform noticeably
better than the random-walk version, especially for the more difficult
cases where $z_\mathrm{out}$ is large.  Given that the random-walk
algorithm is also more computationally demanding, there seems little reason
to use it rather than the shortest-path algorithm, and hence, as discussed
previously, we recommend the latter for most applications.  (To be fair,
the random-walk algorithm does slightly out-perform the shortest-path
algorithm in the example addressed in the following section, although,
being only a single case, it is hard to know whether this is significant.)

\subsection{Zachary's karate club network}
\label{karate}
We now turn to applications of our methods to real-world network data.  Our
first such example is taken from one of the classic studies in social
network analysis.  Over the course of two years in the early 1970s, Wayne
Zachary observed social interactions between the members of a karate club
at an American university~\cite{Zachary77}.  He constructed networks of
ties between members of the club based on their social interactions both
within the club and away from it.  By chance, a dispute arose during the
course of his study between the club's administrator and its principal
karate teacher over whether to raise club fees, and as a result the club
eventually split in two, forming two smaller clubs, centered around the
administrator and the teacher.

\begin{figure}
\begin{center}
\resizebox{\figurewidth}{!}{\includegraphics{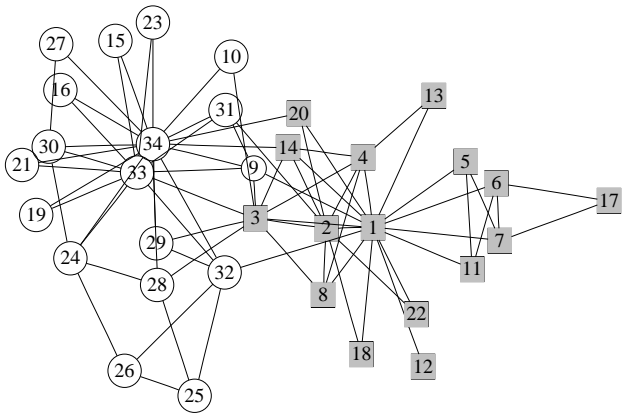}}
\end{center}
\caption{The network of friendships between individuals in the karate club
study of Zachary~\cite{Zachary77}.  The administrator and the instructor
are represented by nodes 1 and 33 respectively.  Shaded squares represent
individuals to who ended up aligning with the club's administrator after
the fission of the club, open circles those who aligned with the
instructor.}
\label{zachnet}
\end{figure}

\begin{figure*}
\begin{center}
\resizebox{17cm}{!}{\includegraphics{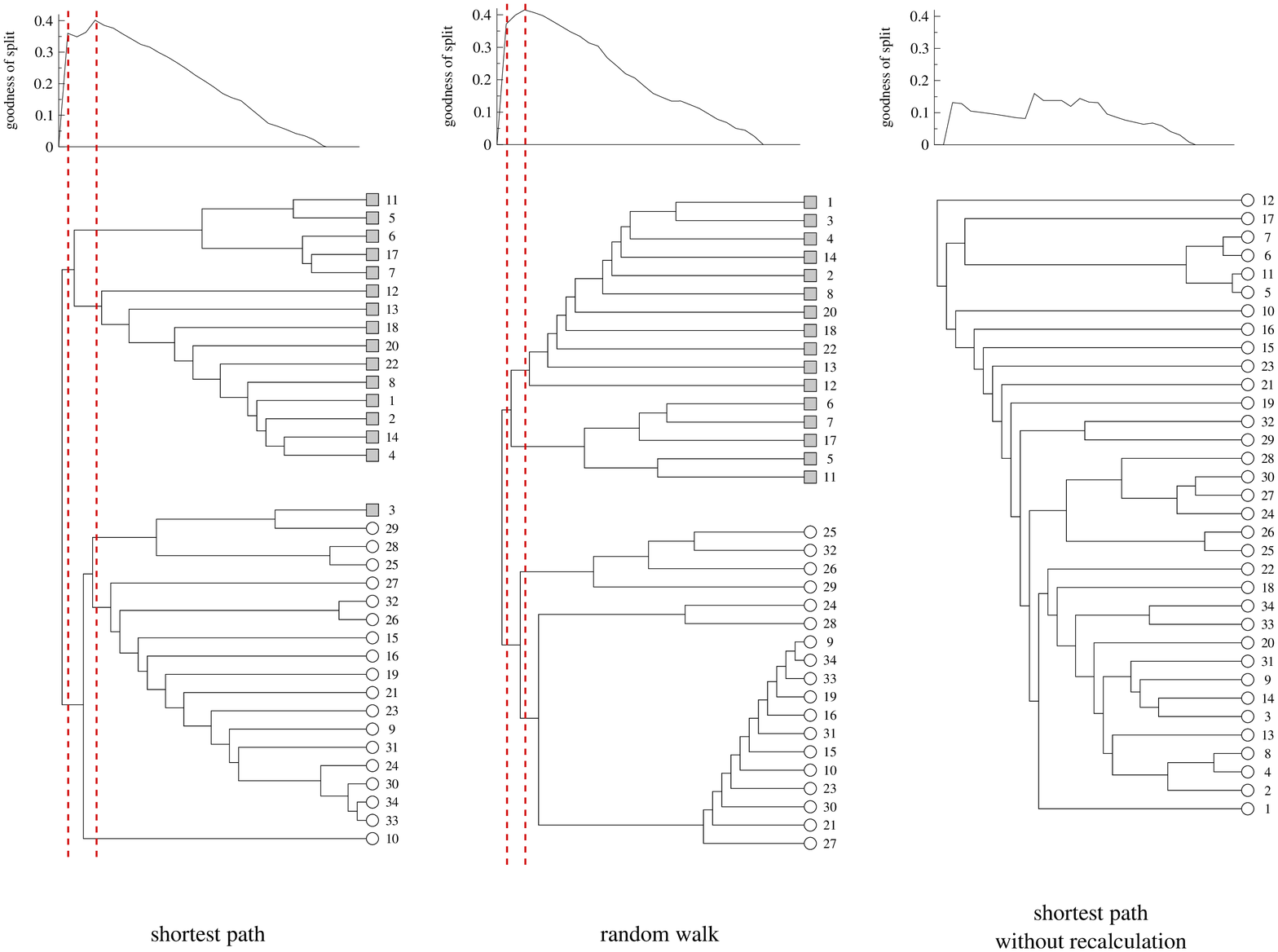}}
\end{center}
\caption{Community structure in the karate club network.   Left: the
dendrogram extracted by the shortest-path betweenness version of our
method, and the resulting modularity.  The modularity has two maxima
(dotted lines) corresponding to splits into two communities (which match
closely the real-world split of the club, as denoted by the shapes of the
vertices) and five communities (though one of those five contains only one
individual).  Only one individual, number~3, is incorrectly classified.
Center: the dendrogram for the random walk version of our method.  This
version classifies all 34 vertices correctly into the factions that they
actually split into (first dotted line), although the split into four
communities gets a higher modularity score (second dotted line).  Right:
the dendrogram for the shortest-path algorithm without recalculation of
betweennesses after each edge removal.  This version of the calculation
fails to find the split into the two factions.}
\label{zachary}
\end{figure*}

In Fig.~\ref{zachnet} we show a consensus network structure extracted from
Zachary's observations before the split.  Feeding this network into our
algorithms we find the results shown in Fig.~\ref{zachary}.  In the
left-most two panels we show the dendrograms generated by the shortest-path
and random-walk versions of our algorithm, along with the modularity
measures for the same.  As we see, both algorithms give reasonably high
values for the modularity when the network is split into two
communities---around 0.4 in each case---indicating that there is a strong
natural division at this level.  What's more, the divisions in question
correspond almost perfectly to the actual divisions in the club revealed by
which group each club member joined after the club split up.  (The shapes
of the vertices representing the two factions are the same as those of
Fig.~\ref{zachnet}.)  Only one vertex, vertex~3, is misclassified by the
shortest-path version of the method, and none are misclassified by the
random-walk version---the latter gets a perfect score on this test.  (On
the other hand, the two-community split fails to produce a local maximum in
the modularity for the random-walk method, unlike the shortest-path method
for which there is a local maximum precisely at this point.)

In the last panel of Fig.~\ref{zachary} we show the dendrogram and
modularity for an algorithm based on shortest-path betweenness but without
the crucial recalculation step discussed in Sec.~\ref{algorithms}.  As the
figure shows, without this step, the algorithm fails to find the division
of the network into the two known groups.  Furthermore, the modularity
doesn't reach nearly such high values as in the first two panels,
indicating that the divisions suggested are much poorer than in the cases
with the recalculation.

\subsection{Collaboration network}
\label{collabnet}
\begin{figure*}
\begin{center}
\resizebox{16.8cm}{!}{\includegraphics{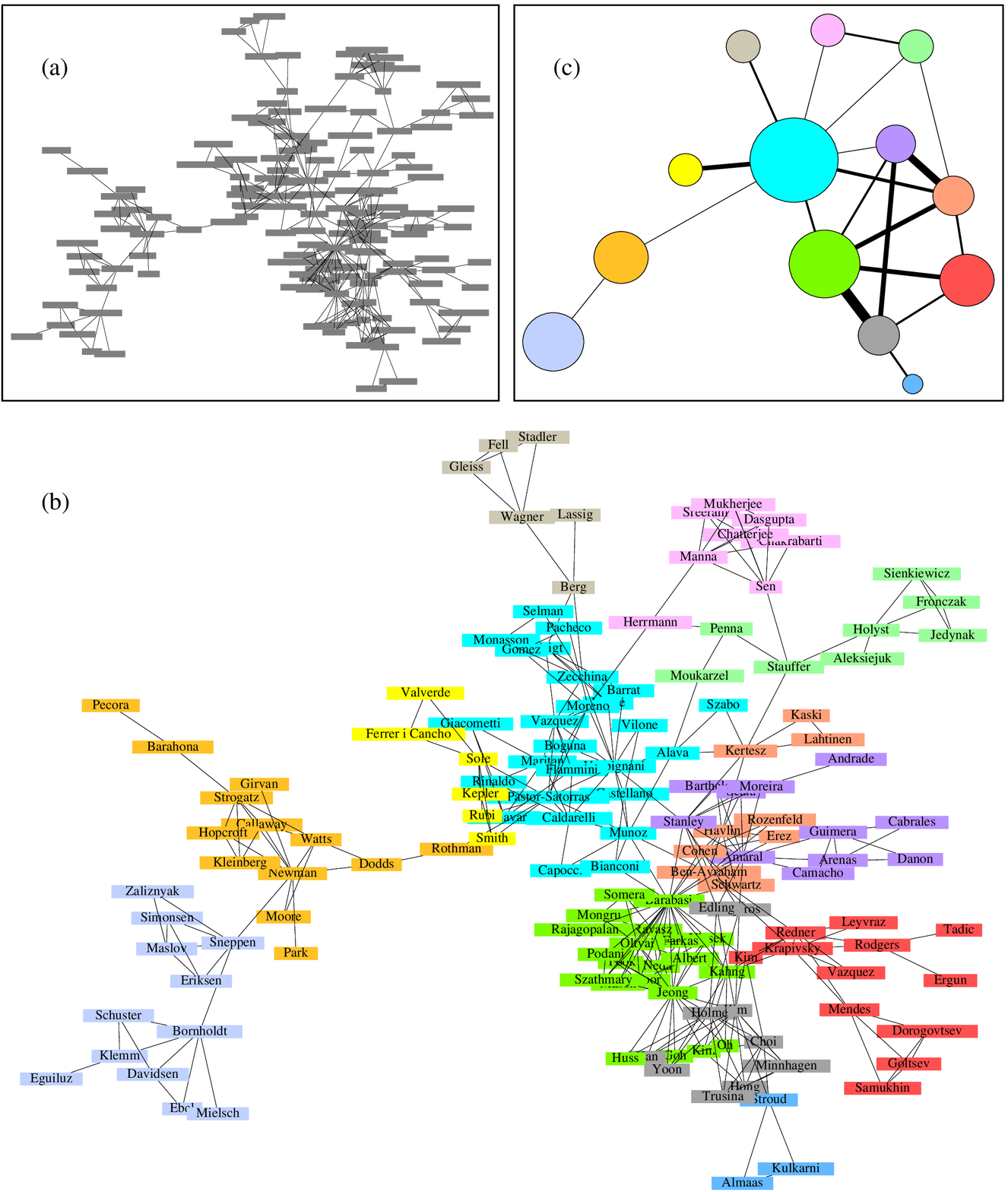}}
\end{center}
\caption{Illustration of the use of the community structure algorithm to
make sense of a complex network.  (a)~The initial network is a network of
coauthorships between physicists who have published on topics related to
networks.  The figure shows only the largest component of the network,
which contains 145 scientists.  There are 90 more scientists in smaller
components, which are not shown.  (b)~Application of the shortest-path
betweenness version of the community structure algorithm produces the
communities shown by the colors.  (c)~A coarse-graining of the network in
which each community is represented by a single node, with edges
representing collaborations between communities.  The thickness of the
edges is proportional to the number of pairs of collaborators between
communities.  Clearly panel~(c) reveals much that is not easily seen in the
original network of panel~(a).}
\label{collab}
\end{figure*}

For our next example, we look at a collaboration network of scientists.
Figure~\ref{collab}a shows the largest component of a network of
collaborations between physicists who conduct research on networks.  (The
authors of the present paper, for instance, are among the nodes in this
network.)  This network (which appeared previously in
Ref.~\onlinecite{PN03}) was constructed by taking names of authors
appearing in the lengthy bibliography of Ref.~\onlinecite{Newman03d} and
cross-referencing with the Physics E-print Archive at
\texttt{arxiv.org}, specifically the condensed matter section of the
archive where, for historical reasons, most papers on networks have
appeared.  Authors appearing in both were added to the network as vertices,
and edges between them indicate coauthorship of one or more papers
appearing in the archive.  Thus the collaborative ties represented in the
figure are not limited to papers on topics concerning networks---we were
interested primarily in whether people know one another, and collaboration
on any topic is a reasonable indicator of acquaintance.

The network as presented in Fig.~\ref{collab}a is difficult to interpret.
Given the names of the scientists, a knowledgeable reader with too much
time on their hands could, no doubt, pick out known groupings, for instance
at particular institutions, from the general confusion.  But were this a
network about which we had no \textit{a priori} knowledge, we would be hard
pressed to understand its underlying structure.

Applying the shortest-path version of our algorithm to this network we find
that the modularity, Eq.~\eref{defsq}, has a strong peak at 13 communities
with a value of $Q=0.72\pm0.02$.  Extracting the communities from the
corresponding dendrogram, we have indicated them with colors in
Fig.~\ref{collab}b.  The knowledgeable reader will again be able to discern
known groups of scientists in this rendering, and more easily now with the
help of the colors.  Still, however, the structure of the network as a
whole and the of the interactions between groups is quite unclear.

In Fig.~\ref{collab}c we have reduced the network to \emph{only} the
groups.  In this panel, we have drawn each group as a circle, with size
varying roughly with the number of individuals in the group.  The lines
between groups indicate collaborations between group members, with the
thickness of the lines varying in proportion to the number of pairs of
scientists who have collaborated.  Now the overall structure of the network
becomes easy to see.  The network is centered around the middle group shown
in cyan (which consists of researchers primarily in southern Europe), with
a knot of inter-community collaborations going on between the groups on the
lower right of the picture (mostly Boston University physicists and their
intellectual descendants).  Other groups (including the authors' own) are
arranged in various attitudes further out.

One of the problems created by the sudden availability in recent years of
large network data sets has been our lack of tools for visualizing their
structure~\cite{Newman03d}.  In the early days of network analysis,
particularly in the social sciences, it was usually enough simply to draw a
picture of a network to see what was going on.  Networks in those days had
ten or twenty nodes, not 140 as here, or several billion as in the world
wide web.  We believe that methods like the one presented here, of using
community structure algorithms to make a meaningful ``coarse graining'' of
a network, thereby reducing its level of complexity to one that can be
interpreted readily by the human eye, will be invaluable in helping us to
understand the large-scale structure of these new network data.

\subsection{Other examples}
\label{other}
In this section, we briefly describe example applications of our methods to
three further networks.  The first is a non-human social network, a network
of dolphins, the second a network of fictional characters, and the third
not a social network at all, but a network of web pages and the links
between them.

\begin{figure}
\begin{center}
\resizebox{\figurewidth}{!}{\includegraphics{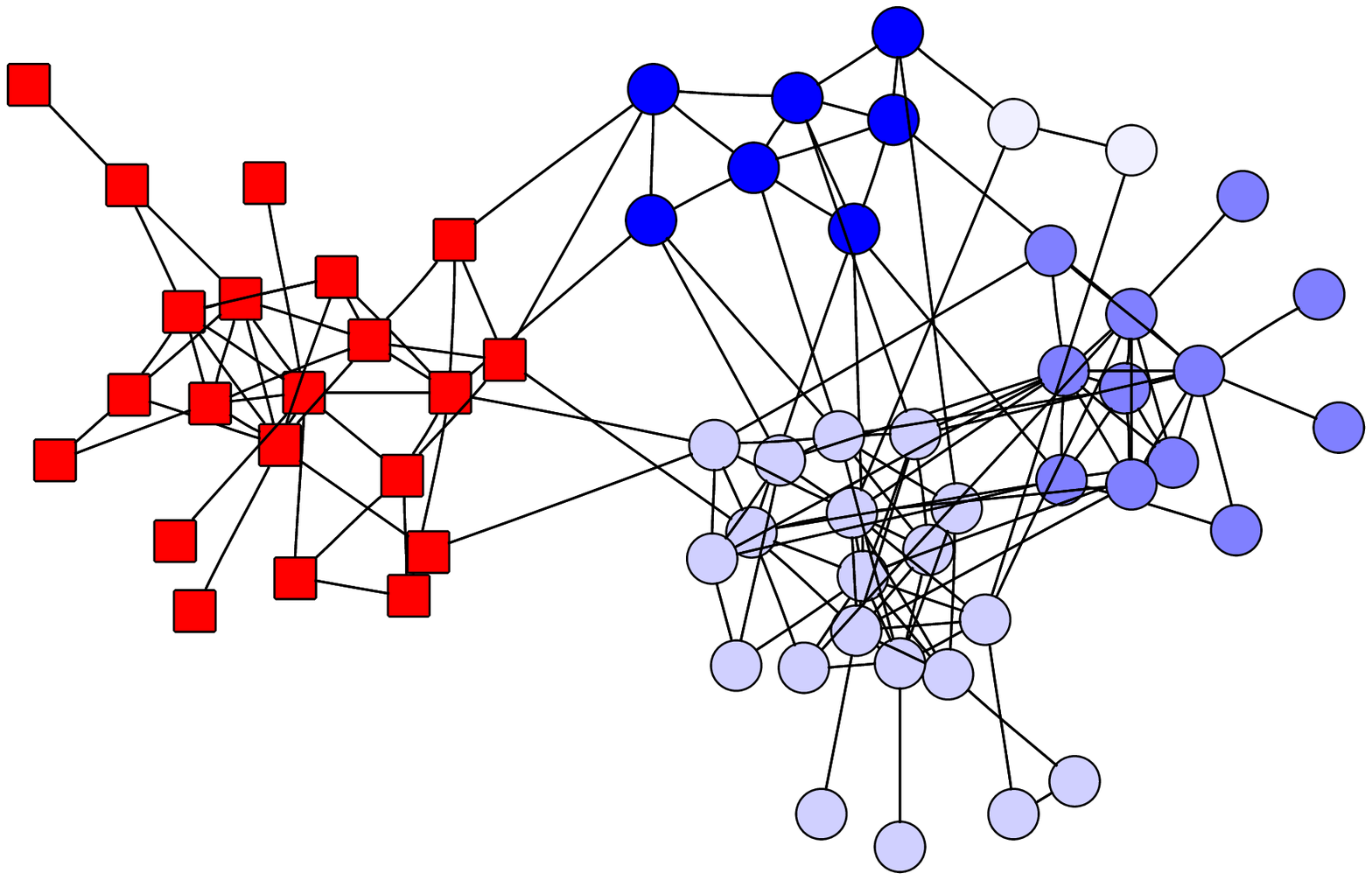}}
\end{center}
\caption{Community structure in the bottlenose dolphins of Doubtful
Sound~\cite{Lusseau03b,Lusseau03a}, extracted using the shortest-path
version of our algorithm.  The squares and circles denote the primary split
of the network into two groups, and the circles are further subdivided into
four smaller groups denoted by the different shades of vertices.  The
modularity for the split is $Q=0.52$.  The network has been drawn with
longer edges between vertices in different communities than between those
in the same community, to make the community groupings clearer.  The same
is also true of Figs.~\ref{lesmis} and~\ref{web}.}
\label{dolphins}
\end{figure}

\begin{figure*}
\begin{center}
\resizebox{14cm}{!}{\includegraphics{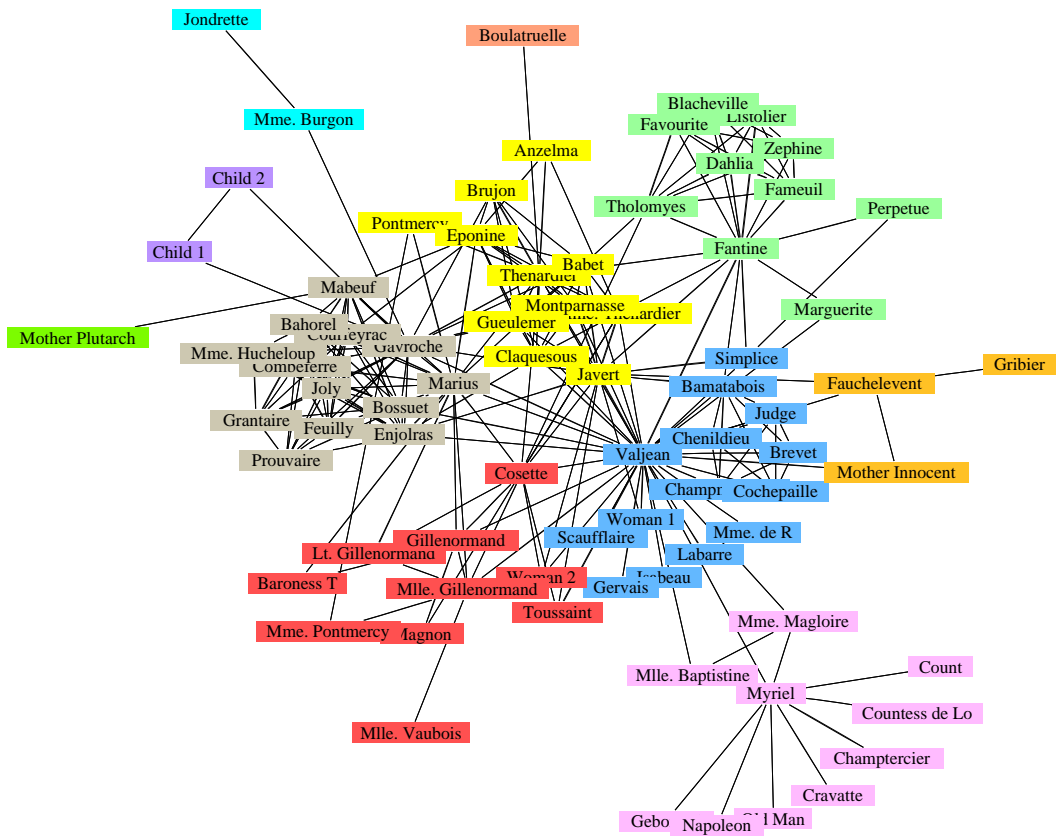}}
\end{center}
\caption{The network of interactions between major characters in the novel
\textit{Les Mis\'erables} by Victor Hugo.  The greatest modularity achieved
in the shortest-path version of our algorithm is $Q=0.54$ and corresponds
to the 11 communities represented by the colors.}
\label{lesmis}
\end{figure*}

\begin{figure}
\begin{center}
\resizebox{\columnwidth}{!}{\includegraphics{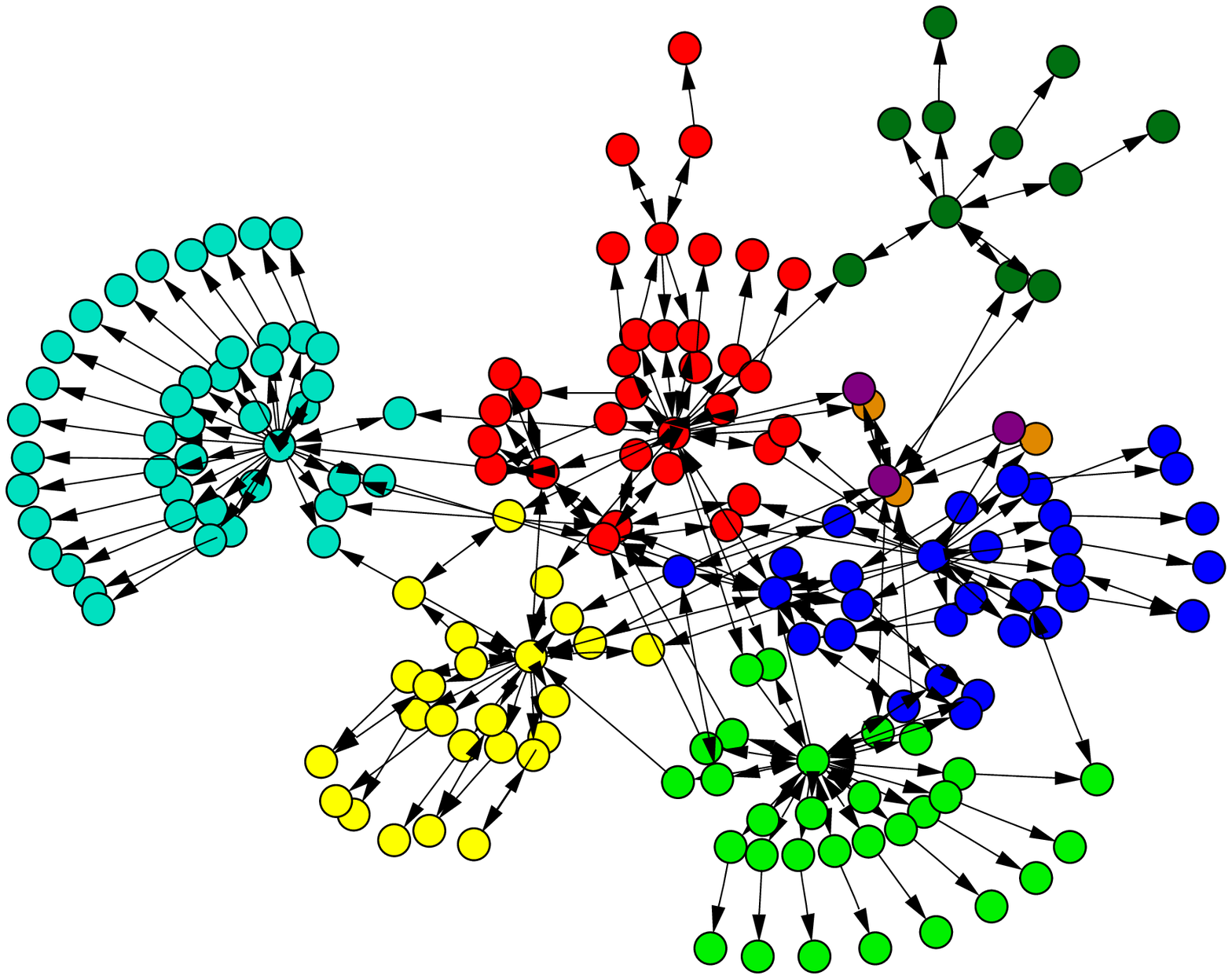}}
\end{center}
\caption{Pages on a web site and the hyperlinks between them.  The colors
denote the optimal division into communities found by the shortest-path
version of our algorithm.}
\label{web}
\end{figure}

In Fig.~\ref{dolphins} we show the social network of a community of 62
bottlenose dolphins living in Doubtful Sound, New Zealand.  The network was
compiled by Lusseau~\cite{Lusseau03b} from seven years of field studies of
the dolphins, with ties between dolphin pairs being established by
observation of statistically significant frequent association.  The network
splits naturally into two large groups, represented by the circles and
squares in the figure, and the larger of the two also splits into four
smaller subgroups, represented by the different shades.  The modularity is
$Q=0.38\pm0.08$ for the split into two groups, and peaks at $0.52\pm0.03$
when the subgroup splitting is included also.

The split into two groups appears to correspond to a known division of the
dolphin community~\cite{Lusseau03a}.  Lusseau reports that for a period of
about two years during observation of the dolphins they separated into two
groups along the lines found by our analysis, apparently because of the
disappearance of individuals on the boundary between the groups.  When some
of these individuals later reappeared, the two halves of the network joined
together once more.  As Lusseau points out, developments of this kind
illustrate that the dolphin network is not merely a scientific curiosity
but, like human social networks, is closely tied to the evolution of the
community.  The subgroupings within the larger half of the network also
seem to correspond to real divisions among the animals: the largest
subgroup consists almost of entirely of females and the others almost
entirely of males, and it is conjectured that the split between the male
groups is governed by matrilineage (D.~Lusseau, personal communication).

Figure~\ref{lesmis} shows the community structure of the network of
interactions between major characters in Victor Hugo's sprawling novel of
crime and redemption in post-restoration France, \textit{Les Mis\'erables}.
Using the list of character appearances by scene compiled by
Knuth~\cite{Knuth93}, the network was constructed in which the vertices
represent characters and an edge between two vertices represents
co-appearance of the corresponding characters in one or more scenes.  The
optimal community split of the resulting graph has a strong modularity of
$Q=0.54\pm0.02$, and gives 11 communities as shown in the figure.  The
communities clearly reflect the subplot structure of the book:
unsurprisingly, the protagonist Jean Valjean and his nemesis, the police
officer Javert, are central to the network and form the hubs of communities
composed of their respective adherents.  Other subplots centered on Marius,
Cosette, Fantine, and the bishop Myriel are also picked out.

Finally, as an example of the application of our method to a non-social
network, we have looked at a web graph---a network in which the vertices
and edges represent web pages and the links between them.  The graph in
question represents 180 pages from the web site of a large
corporation~\footnote{The graph is one of the test graphs from the graph
drawing competition held in conjunction with the Symposium on Graph
Drawing, Berkeley, California, September 18--20, 1996.}.  Figure~\ref{web}
shows the network and the communities found in it by the shortest-path
version of our algorithm.  This network has one of the strongest modularity
values of the examples studied here, at $Q=0.65\pm0.02$.  The links between
web pages are directed, as indicated by the arrows in the figure, but, as
discussed in Sec.~\ref{eb}, for the purposes of finding the communities we
ignore direction and treat the network as undirected.

Certainly it might be useful to know the communities in a web network; an
algorithm that can pick out communities could reveal which pages cover
related topics or the social structure of links between pages maintained by
different individuals.  Ideas along these lines have been pursued by, for
example, Flake~\etal~\cite{FLGC02} and Adamic and Adar~\cite{AA01}.

\section{Conclusions}
\label{concs}
In this paper we have described a new class of algorithms for performing
network clustering, the task of extracting the natural community structure
from networks of vertices and edges.  This is a problem long studied in
computer science, applied mathematics, and the social sciences, but it has
lacked a satisfactory solution.  We believe the methods described here give
such a solution.  They are simple, intuitive, and demonstrably give
excellent results on networks for which we know the community structure
ahead of time.  Our methods are defined by two crucial features.  First, we
use a ``divisive'' technique which iteratively removes edges from the
network, thereby breaking it up in communities.  The edges to be removed
are identified by using one of a set of edge betweenness measures, of which
the simplest is a generalization to edges of the standard shortest-path
betweenness of Freeman.  Second, our algorithms include a recalculation
step in which betweenness scores are re-evaluated after the removal of
every edge.  This step, which was missing from previous algorithms, turns
out to be of primary importance to the success of ours.  Without it, the
algorithms fail miserably at even the simplest clustering tasks.

We have demonstrated the efficacy and utility of our methods with a number
of examples.  We have shown that our algorithms can reliably and
sensitively extract community structure from artificially generated
networks with known communities.  We have also applied them to real-world
networks with known community structure and again they extract that
structure without difficulty.  And we have given examples of how our
algorithms can be used to analyze networks whose structure is otherwise
difficult to comprehend.  The networks studied include a collaboration
network of scientists, in which our methods allow us to generate schematic
depictions of the overall structure of the network and collaborations
taking place within and between communities, other social networks of
people and of animals, and a network of links between pages on a corporate
web site.

The primary remaining difficulty with our algorithms is the relatively high
computational demands they make.  The fastest of them, the one based on
shortest-path betweenness, operates in $\O(n^3)$ time on a sparse graph,
which makes it usable for networks up to about $10\,000$ vertices, but for
larger systems it becomes intractable.  Although the ever-improving speed
of computers will certainly raise this limit in coming years, it would be
more satisfactory if a faster version of the method could be discovered.
One possibility is parallelization: the betweenness calculation involves a
sum over source vertices and the elements of that sum can be distributed
over different processors, making the calculation trivially parallelizable
on a distributed-memory machine.  However, a better approach would be to
find some improvement in the algorithm itself to decrease its computational
complexity.

Since the publication of our first paper on this topic~\cite{GN02}, several
other authors have made use of the shortest-path version of our algorithm.
Holme~\etal~\cite{HHJ02} have applied it to a number of metabolic networks
for different organisms, finding communities that correspond to functional
units within the networks, while Wilkinson and Huberman~\cite{WH03} have
applied it to a network of relations between genes, as established by
co-occurrence of names of genes in published research articles.  An
interesting application to social networks is the study by Gleiser and
Danon~\cite{GD04} of the collaboration network of early jazz musicians.
They found, among other things, that the network split into two communities
along lines of race, black musicians in one group, white musicians in the
other.  Guimer\`a~\etal~\cite{Guimera03} have applied the method to a
network of email messages passing between users at a university, and found
communities that reflect both formal and informal levels of organization.
Tyler~\etal~\cite{TWH03} have also applied the algorithm to an email
network, in their case at a large company, finding that the resulting
communities correspond closely to organizational units.  The latter work is
interesting also in that it suggests a method for improving the speed of
the algorithm: Tyler~\etal\ calculate betweenness for only a subset,
randomly chosen, of possible source vertices in the network, rather than
summing over all sources.  The size of the subset is decided on the fly, by
sampling source vertices until the betweenness of at least one edge in the
network exceeds a predetermined threshold.  This technique reduces the
running time of the calculation considerably, although the resulting
estimate of betweenness necessarily suffers from the statistical
fluctuations inherent in random sampling methods.  This idea, or a
variation of it, might provide a solution to the problems mentioned above
of the high computational demands of our algorithms.

We are of course delighted to see our methods applied to such a variety of
problems.  Combined with the new algorithms and measures described in this
paper, we hope to see many more applications in the future.

\begin{acknowledgments}
The authors thank Steven Borgatti, Ulrik Brandes, Linton Freeman, David
Lusseau, Mason Porter, and Douglas White for useful comments.  Thanks also
to Oliver Boisseau, Patti Haase, David Lusseau, and Karsten Schneider for
providing the data for the dolphin network and to Douglas White for the
karate club data.  This work was funded in part by the National Science
Foundation under grant number DMS--0234188 and by the Santa Fe Institute.
\end{acknowledgments}

\end{document}